\documentstyle[aps,multicol,epsf,epsfig,axodraw]{revtex}

\begin{document}
%\draft
%\twocolumn[\hsize\textwidth\columnwidth\hsize\csname @twocolumnfalse\endcsname

\title{Polyelectrolyte Solutions with Multivalent Salts}
\author{Paulo S. Kuhn (1) , Yan Levin (1) \footnote{Corresponding author}, and Marcia C. Barbosa (1,2) \\
(1) Instituto de F\'{\i}sica, Universidade Federal
do Rio Grande do Sul\\ Caixa Postal 15051, CEP 91501-970
, Porto Alegre, RS, Brazil\\
(2) Institute for Theoretical Physics \\ University of California, Santa Barbara,
CA, 93106-2431, USA\\
e-mails: kuhn@if.ufrgs.br, levin@if.ufrgs.br,\\ barbosa@if.ufrgs.br, 
barbosa@itp.ucsb.edu}
\maketitle
\begin{abstract}
We investigate the thermodynamic properties of a polyelectrolyte solution in a presence of {\it multivalent} salts.   The polyions are modeled
as rigid cylinders with the charge distributed uniformly along the major axis. The solution, besides the polyions, contain monovalent and divalent counterions 
as well as monovalent coions. The strong electrostatic
attraction existing between the polyions and the counterions results in formation
of clusters consisting of one polyion and a number of associated monovalent
and divalent counterions.   The theory presented in the paper allows us to 
explicitly construct the Helmholtz free energy of a polyelectrolyte solution.  The characteristic cluster size, as well as any other thermodynamic property can then be determined by an appropriate operation on the free energy.
\end{abstract}
\pacs{PACS numbers:05.70.Ce; 61.20.Qg; 61.25.Hq}

%]
\bigskip

\begin{multicols}{2}

\centerline{\bf 1. Introduction}

\bigskip

The polyelectrolytes are ubiquitous in our every day life.  They are 
present in a variety of industrial products ranging from the hair conditioners
to the superabsorbants used in baby diapers.  The carrier of genetic code, the DNA, is also
a polyelectrolyte.  Not withstanding their omnipresence, our understanding
of this class of polymers is far from complete.  The long range Coulomb interaction
combined with a strong asymmetry existing between the various entities of a polyelectrolyte solution, has proven to be a formidable challenge to the standard methods of liquid state theory.  The scaling approaches which were so successful
in elucidating the properties of non-ionic polymers have, so far, failed in their
extension to polyelectrolytes.  What
seems to be lacking is a mean-field theory similar to the one constructed
by Flory for the simple polymers and by Debye and H\"uckel for the symmetric electrolytes.
Recently, we have attempted to construct such a mean-field theory for the 
special class of rigid
polyelectrolytes, an example of which is provided by a solution of DNA segments.
The advantage of working with a rod-like polyelectrolytes is that the intramolecular
degrees of freedom associated with the conformational state of a polyion can be neglected.  The polyelectrolyte solution then becomes a strongly asymmetric
electrolyte, to which a generalization of the Debye-H\"uckel-Bjerrum (DHBj)  theory \cite{Fi93} can be applied.  Additional advantage in working with rigid
polyelectrolytes is that they obey the Manning infinite dilution 
limiting laws \cite{Ma69}.  These laws are extremely useful since they provide the boundary conditions that any theory of rod-like polyelectrolytes must
satisfy.  

We have demonstrated that the extension of the DHBj theory to rigid polyelectrolytes, indeed, satisfies the Manning Limiting Laws \cite{Le96}.  
Furthermore, the theory is in excellent
agreement with the experiments on rigid polyelectrolyte solutions in a presence of monovalent salt \cite{Ku98}.  Recently we have extended our work to study the interaction of a polyelectrolyte  with a cationic surfactants \cite{Ku298}.  The latter problem is of great practical interest in a variety of biochemical and biomedical applications related to design of gene delivery systems \cite{Fe89}.  Again,
the DHBj theory has proven successful in giving quantitative agreement with experiments without a need for any adjustable parameters.  In the current work
we shall extend the theory to allow for the presence of both the monovalent and the divalent salts.  
This problem, once again, is motivated by the biomedical applications to gene
therapy.  It has been recently found that introduction of a DNA into host cells
is facilitated by the presence of a small amount of divalent salt such as calcium phosphate \cite{Fe89}.  

Since the bulk of the calculations is similar to those presented in the earlier
work, we shall not enter in great detail and instead refer the reader to the
previously published papers \cite{Le96,Ku98,Ku298}.

\begin{figure}[h]

%\vspace*{-3cm}
%\vspace*{-4cm}
\begin{center}
\epsfxsize=8.cm
\leavevmode\epsfbox{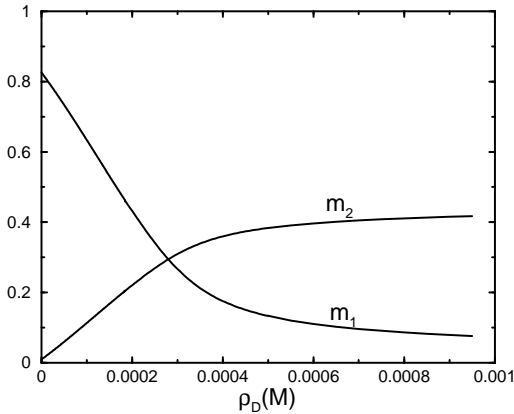}
\end{center}
%\vspace*{-0.5cm}
\vspace*{0.5cm}
\begin{minipage}{0.48\textwidth}
\caption{The DNA binding isotherms for the monovalent and divalent counterions. The parameters are: $Z=440$, $\xi=4.17$, $\rho_p=2 \times 10^{-6}$M, $\rho_M=10^{-2}$M, $a_p=27\AA$, $a_c=7.04\AA$.}
\label{Fig.1}
\end{minipage}
\end{figure}
\noindent

\bigskip
\bigskip
 
\centerline{\bf 2. The model}

\bigskip
\bigskip

We begin by  defining the Primitive Model of Polyelectrolyte 
(PMP) \cite{Le96}. Our system consist of long cylindric polyions, 
spherical counterions, monovalent salt, and divalent salt, inside the
volume $V$. The polyions, of length $L$, have diameter $a_p$ and charge $-Zq$ uniformly distributed along the length of the cylinder. The 
distance between the charged groups is $b \equiv L/Z$. The counterions 
and coions are modeled as rigid spheres of diameter $a_c$ and 
charge $+q$,  $+2q$, $-q$  located at their centers.  We shall not distinguish 
between the monovalent counterions derived from the polymers or salt, 
and assume that all the salt is dissociated.  Similarly, to simplify calculations,
we shall treat the coions derived from the monovalent and the divalent salts as being identical. 
The solvent is modeled as a continuum medium of dielectric constant $D$. 
The distance of closest approach, $a$, between a polyion and a counterion is 
$a = (a_p+a_c)/2$. This represents the ``exclusion cylinder" due to 
the hard-core repulsion. 

The strong electrostatic attraction existing 
between the macromolecules and the counterions 
 results in some number of counterions becoming associated with the 
polyions, producing clusters composed of {\it one} 
polyion, $m_B$ monovalent counterions, and $d_B$ 
divalent counterions. In fact, we expect that there shall be 
a distribution of cluster sizes. Motivated, however, by our 
previous work \cite{Ku98} we know that the polydispersity in 
cluster sizes is not very important, if one is interested in thermodynamics of a polyelectrolyte solution \cite{Le98}. 
Since not all of 
the counterions condense onto the polyions, for any non-zero 
temperature some free, unassociated, counterions remain in the 
solution. We denote the density of unassociated monovalent 
counterions as $\rho_+= (Z-m_B) \rho_p + \rho_M$, and the density of
 unassociated divalent counterions as
$\rho_{++}=\rho_D - d_B \rho_p$, where $\rho_p$, $\rho_M$, and $\rho_D$ are the densities of the polyions, the monovalent salt, and the divalent salt respectively.
Since the coions do not participate in association their density remains 
unchanged, $\rho_-=\rho_M+2 \rho_D$.  The goal of the theory 
is to determine the characteristic cluster size, 
i.e. the values of $m_B$ and $d_B$. 

Complete thermodynamic information about the system 
is contained in its Helmholtz free energy. The 
condition that the free energy must be minimum  
allows us to determine the values of $m_B$ and $d_B$. Once 
this is done, all the thermodynamic functions of the system 
can be found through the appropriate operations on the free 
energy. For example, the pressure inside the 
polyelectrolyte solution is a Legendre transform of the 
Helmholtz free energy density $f=-F/V$, given by
$p(T,\{\rho_t\}) = f(T,\{\rho_t\}) + \sum \mu_t \rho_t$,
where the chemical potential of a specie 
of type $t$ (clusters, counterions, and coions) is 
$\mu_t=-\partial f/ \partial \rho_t$.

The free energy 
cannot be calculated 
exactly. We shall, therefore, attempt 
to construct it as 
a sum of the most relevant contributions. These 
can be divided into an electrostatic and an 
entropic ones. The electrostatic contribution 
arises as the result of polyion-counterion, polyion-polyion, and 
the counterion-coion interactions. The entropic 
contribution is the result of mixing of various species \cite{Le96}.

The free energy for the 
polyion-counterion interaction is 
calculated in the framework of  the
Debye-H\"uckel (DH) theory \cite{DH23}. The full 
calculation is presented in Ref.\cite{Ku98} and here we shall
just quote the result, 

\begin{eqnarray}
\beta f^{pc} & = & \rho_p (Z-m_B-2 d_B)^2 
\frac {(a/L)} {T^* (\kappa a)^2} \times \nonumber \\
& & \times \left\{ 2 \ln \left[ \kappa a K_1(\kappa a) \right] 
- I(\kappa a) + \frac {(\kappa a)^2} {2} \right\} 
\end{eqnarray}
with $\kappa a=\sqrt(4\pi\rho_1^{\ast}/T^{\ast})$, 
$T^{\ast}=Dk_BT a/q^2$ is the reduced temperature,  
$\rho_1^{\ast}=\rho_1 a^3$, and

\begin{equation}
I(\kappa a) \equiv \int^{\kappa a}_{0} dx 
\frac {x K_0^2(x)} {K_1^2(x)} \, .
\end{equation}

\noindent
where $K_n$ is the Bessel function of order $n$.
The modification of replacing the total microion density by {\it twice}
the ionic strength, $\rho_1 \equiv \rho_+ + 4 \rho_{++} + \rho_-  $, is necessary when discussing multivalent salts.

The contribution to the free 
energy due to the interactions among the various 
free ions is given by the usual DH theory \cite{DH23},

\begin{equation}
\label{e22}
\beta f^{cc} = \frac {1} {4 \pi} 
\frac {1} {a_c^3} \left\{ \ln 
\left( 1+\kappa a_c \right) - \kappa a_c 
+ \frac {(\kappa a_c)^2} {2} \right\} \, .
\end{equation} 

The free, unassociated, counterions and coions screen the polyion-polyion
 interactions,
producing an effective potential 
of short range. The polyion-polyion 
contribution to the free energy can then be calculated using a Van der 
Waals type of approach \cite{Ku98}. To this end the 
polyion-polyion contribution is expressed as a 
second virial term, averaged over the relative angle 
sustained by two molecules. The result is,

\begin{equation}
\label{e23}
\beta f^{pp} = \frac {-2 \pi a^3 \exp (-2 \kappa a) 
(Z-m_B-2d_B)^2 {\rho_p}^2} {T^* (\kappa a)^4 K_1(\kappa a)^2} \, .
\end{equation} 

The entropic contribution is found using the Flory theory \cite{Fl67}. Thus, 
the increase in free energy due to mixing of various 
species is the sum of their ideal free energies,

\begin{equation}
\label{e1}
\beta f^{ent} = \sum_t \left[ \rho_t - \rho_t \ln \left( \phi_t/\zeta_t \right) \right] \, ,
\end{equation}

\noindent
where, in obvious notation, the respective volume fractions are,

\begin{eqnarray}
\label{e2}
& & \phi_{cl} = \frac {\pi \rho^*_p} {4 (a/L)} \left(\frac {a_p} {a}\right)^2 + (m_B+d_B) \frac {\pi \rho^*_p} {6} \left(\frac {a_c} {a}\right)^3 \, , \\
& & \phi_+ = \frac {\pi \rho^*_+} {6} \left( \frac {a_c} {a} \right)^3 \, , \\
& & \phi_{++} = \frac {\pi \rho^*_{++}} {6} \left( \frac {a_c} {a} \right)^3 \, , \\
& & \phi_- = \frac {\pi \rho^*_-} {6} \left( \frac {a_c} {a} \right)^3 \, .
\end{eqnarray}

\noindent
 $\zeta_t$ is the internal 
partition function of an isolated specie $t$. For structureless 
particles, $\zeta_+=\zeta_-=\zeta_{++} \equiv 1$. The calculation of $\zeta_{cl}$, 
proceeds along the lines given in Ref. \cite{Ku98,Ku298}. We find that to a good accuracy the internal partition function of a cluster can be approximated by

\begin{eqnarray}
\label{e8a}
\ln \zeta_{cl} & = & -\xi \, S \, (-2m_1 - 4m_2    +4m_1m_2 +m_1^2  +4m_2^2) + \nonumber \\
& & - Z \,  [ (1-m_1-m_2) \ln \left( 1-m_1-m_2 \right) + \nonumber \\
& & + m_1 \ln m_1 + m_2 \ln m_2 ] \, ,
\end{eqnarray}

\noindent
where the binding fractions are, $m_1=m_B/Z$, $m_2=d_B/Z$,  and $S$ is 

\begin{equation}
\label{e9}
S \equiv \frac {1} {2} \sum_{s_1 \neq s_2} \frac {1} {|s_1-s_2|} = Z [\psi(Z) - \psi(1)] -Z +1 \, ,
\end{equation}

\noindent
where $\psi(n)$ is the digamma function  
and $\xi \equiv \beta q^2/Db$ is the Manning parameter \cite{Ma69}.

\bigskip
\bigskip

\centerline{ \bf 3. Results and Conclusions}

\bigskip
\bigskip

Adding the above contributions
we obtain the
full Helmholtz free energy for the polyelectrolyte plus 
salt system,
$f = f^{ent} + f^{pc} + f^{cc} + f^{pp} $.
 Minimization of the free energy allows us to find the number of condensed monovalent and divalent counterions, 
 $m_B$ and $d_B$. In 
Fig.$1$ we show the dependence of the binding fractions on the 
density of divalent salt, at fixed 
polyion and monovalent 
salt concentrations. At low densities of divalent salt, the condensation
is dominated by the monovalent ions,
since the divalent ions gain more entropy, and thus lower the total free energy, by staying free. However as the 
concentration of divalent salt increases, the gain in electrostatic energy due
to condensation wins over the entropy and more divalent ions become associated with the macromolecules. We note that unlike the association with the ionic surfactants, which exhibits a large degree of cooperativity characterized by the sharp rise in the surfactant binding fraction, the replacement of the condensed monovalent counterions by the divalent counterions proceeds quite smoothly.

\begin{figure}[h]

%\vspace*{-3cm}
%\vspace*{-4cm}
\begin{center}
\epsfxsize=8.cm
\leavevmode\epsfbox{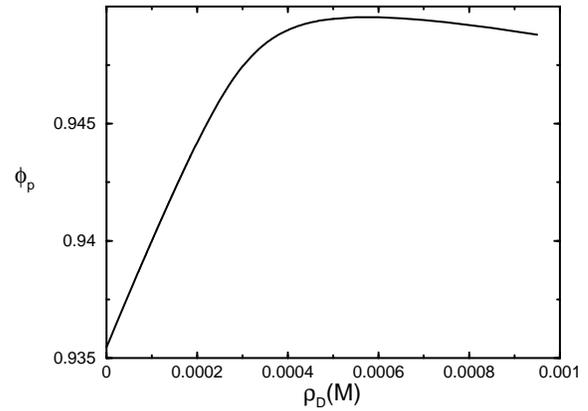}
\end{center}
%\vspace*{-0.5cm}
\vspace*{-0.25cm}
\begin{minipage}{0.48\textwidth}
\caption{The osmotic pressure coefficient for the polyelectrolyte multivalent salt system. The parameters have the same value as in Fig.1.}
\label{Fig.2}
\end{minipage}
\end{figure}
\noindent

This result could have been anticipated {\it a priori}.  The presence of a condensed divalent
counterion does not stimulate any further condensation of additional particles, 
but on
the contrary, the electrostatic repulsion between the like-charged counterions tends to inhibit any further association.  The situation is quite different in the case of binding by the ionic surfactants.  After the first amphiphile is associated,
the condensation of additional molecules is energetically favored since the buildup of the hydrocarbon density in the vicinity of a polyion helps to exclude water and, thus, reduces the unfavorable hydrophobic energy of the alkyl tails.  This, then, explains
why the association with ionic surfactant and lipids exhibits a large degree of cooperativity while the binding of divalent counterions proceeds in a completely uncooperative fashion.  

Finally, using the number of associated ions, the
osmotic pressure, defined as $\beta p \equiv (Z \rho_p+2\rho_M+3 \rho_D) \phi_p$, can be computed   
as a function of divalent salt density (Fig. $2$). 

\section*{\it ACKNOWLEDGMENTS}

This work was supported in part by CNPq - Conselho 
Nacional de
Desenvolvimento Cient\'{\i}fico e Tecnol\'ogico 
and FINEP -
Financiadora de Estudos e Projetos, Brazil. This research
was also supported by the National Science Foundation
under Grant No. PHY94-07194.

%\newpage

\end{multicols}

\begin{thebibliography}{40}

\bibitem {Fi93} M. E. Fisher and Y. 
Levin, {\it Phys. Rev. Lett.} {\bf 71}, 3826 
(1993); Y.Levin and M. E. Fisher, {\it Physica A} {\bf 225}, 164 (1996).

\bibitem {Ma69} G.S. Manning, {\it Quart. Rev. of Biophys. II} {\bf 2}, 179 (1978); J.L. Barrat and J. F. Joanny, {\it Adv. Chem. Phys.}
{\bf 94}, 1, (1996); G. S. Manning, {\it J. Chem. Phys.} {\bf 51}, 924 (1969).

\bibitem {Le96} Y. Levin, {\it Europhys. Lett.} {\bf 34}, 405 (1996); Y. 
Levin 
and M. C. Barbosa, {\it J. Phys. II (France)} {\bf 7}, 37 (1997).

\bibitem {Ku98} P. S. Kuhn, Y. Levin, and M. C. Barbosa, 
{\it Macromolecules} (in press).

\bibitem {Ku298} P. S. Kuhn, Y. Levin, and M. C. Barbosa, "Complex Formation
Between Polyelectrolyte and Ionic Surfactant" con-mat/9806226.

\bibitem{Fe89} P. L. Felgner and G. M. Ringold, {\it Nature} {\bf 337}, 387 (1989)

\bibitem {Le98} Y. Levin, M. C. Barbosa, and M. N. Tamashiro {\it Europhys. Lett.} {\bf 41}, 123 (1998)

\bibitem {DH23} (a) P.W. Debye and E. H\"uckel, {\it Phys. Z.} {\bf 24}, 
185 (1923);
(b) For a good exposition see D.A. McQuarrie, {\it Statistical 
Mechanics} (Harper and Row, New York, 1976), Chap. 15.

\bibitem {Fl67} P. Flory, {\it Principles of 
Polymer Chemistry} (Cornell University Press, Ithaca, New York, 1971).

\end{thebibliography}
\end{document}